\documentclass[12pt]{article}
\usepackage{amsmath}
\usepackage{amssymb,epsfig}
\newcommand{\bea}{\begin{eqnarray}}	
\newcommand{\eea}{\end{eqnarray}}

\newtheorem{theorem}{Theorem}

\begin{document}

\title{The Ponzano-Regge asymptotic of the $6j$ symbol: an elementary proof}
\author{Razvan Gurau \footnote{e-mail: rgurau@perimeterinstitute.ca}\\
Perimeter Institute, 31 Caroline St. N,\\ Waterloo, ON N2L 2Y5, Canada.}
\maketitle

\begin{abstract}
In this paper we give a direct proof of the Ponzano-Regge asymptotic formula for the Wigner $6j$ symbol starting from 
Racah's single sum formula. Our method treats halfinteger and integer spins on the same footing. The generalization to 
Minkowskian tetrahedra is direct. This result should be relevant for the introduction of renormalization scales in 
spin foam models.
\end{abstract}

\section{Introduction}

The connection between the renormalization group, so successful in describing low energy physics, and the 
theory of loop quantum gravity is still an open question. A promissing line of research is to explore in detail
the relationship between the renormalization group and the spin foam quantization of gravity \cite{Perez:2004hj} (more precisely the group field theory (GFT) dual to spin foams).

GFT \cite{FreidelGFT} can be represented either in terms of group integrals or in terms of tensor models.
The situation is highly reminiscent of the one encountered in noncommutative quantum field theory (NCQFT) \cite{Douglas:2001ba,Connes:1997cr}, 
where the group integral formulation is similar to the direct space representation \cite{Gurau:2005gd} and 
the tensorial model is similar to the matrix base representation \cite{Rivasseau:2005bh,Disertori:2006nq}.

Particularly, the dual GFT of $3D$ quantum gravity resembles a $\phi^4$ model. We can describe it in terms of a 
tensorial field theory, with a vertex weight given by  Wigner's $6j$ symbol, and a trivial propagator \cite{FreidelGFT}. 
This setting, although very encouraging, is not yet adapted to renormalization.
The triviality of the propagator makes the definition of scales 
unclear. The parallel problem in the NCQFT has been solved by the introduction of spectral scales, first in the matrix base
\cite{Rivasseau:2005bh}, and then in the direct space \cite{Gurau:2005gd}. Consequently, one may argue that the scales in GFT could be more readily accessible in the tensor model formulation. A deeper understanding of this model is then required.

It is well known that the $6j$ symbol obeys the Ponzano-Regge asymptotic formula \cite{PR} . Several proofs of this formula exist
\cite{Roberts,Freidel:2002mj}, but they rely either on an algebraic definition or on a group integral definition of the $6j$ symbol.

The goal of this paper is to present an alternative proof of this asymptotic formula, developed entirely in the discrete space of indexes of the tensor model, hence, in a formalism presumably better suited to the introduction of scales and renormalization.

In the next section we give some notations and state our main theorem. The following section consists of its proof. The last section elaborates on the different generalizations of our method.

\section{Notations and Main Theorem}
There are several ways of expressing the $6j$ symbol. The starting point of our derivation is Racah's single sum formula
\bea\label{Racah}
\genfrac{\{}{\}}{0pt}{}{j_1\;j_2\;j_3}{J_1\;J_2\;J_3}=&&\sqrt{\Delta(j_1,j_2,j_3)\Delta(J_1,j_2,J_3)\Delta(J_1,J_2,j_3)\Delta(j_1,J_2,J_3)}\nonumber\\
&&\sum_{\text{max } v_i}^{\text{min }p_j}(-1)^t\frac{(t+1)!}{\prod_i(t-v_i)!\prod_j(p_j-t)!} \; ,
\eea
with
\bea\label{eq:vp}
&&v_1=j_1+j_2+j_3 \quad v_2=J_1+j_2+J_3 \nonumber\\
&&v_3=J_1+J_2+j_3 \quad v_4=j_1+J_2+J_3\nonumber\\
&&p_1=j_2+J_2+j_3+J_3 \quad p_2=j_1+J_1+j_3+J_3 \quad p_3=j_2+J_2+j_1+J_1 \nonumber\\
&&\Delta(j_1,j_2,j_3)=\frac{(j_1+j_2-j_3)!(j_1-j_2+j_3)!(-j_1+j_2+j_3)!}{(j_1+j_2+j_3+1)!} \; .
\eea

The $j$'s and $J$'s are integers or halfintegers. The sum in eq. (\ref{Racah}) is over all integers $t$ such that all the arguments of the factorials are positive. The $\Delta(j_1,j_2,j_3)$ factors are called triangle coefficients.

\begin{figure}[hbt]
\centerline{
\includegraphics[width=50mm]{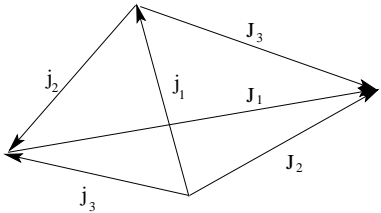}
}
\caption{Labeling of the tetrahedron}\label{fig:tet}
\end{figure}

This weight is associated with an euclidean tetrahedron with edges $j_1$, $j_2$, $j_3$, $J_1$, $J_2$, $J_3$ labeled as in figure \ref{fig:tet}.

We can rewrite $j_2, J_3, J_1$ in terms of $j_1,J_2,j_3$ like
\bea\label{mese}
\vec{j_2}=\vec{j_3}-\vec{j_1} \quad \vec{J_3}=\vec{J_2}-\vec{j_1} \quad \vec{J_1}=\vec{J_2}-\vec{j_3} \; .
\eea 

We will prove the following asymptotic formula for the $6j$ symbol

\begin{theorem} \label{theo}
 Under a rescaling of all its arguments by a large $k$ the $6j$ symbol behaves like
\bea
&&\genfrac{\{}{\}}{0pt}{}{kj_1\;kj_2\;kj_3}{kJ_1\;kJ_2\;kJ_3}\nonumber\\
&&=\frac{1}{\sqrt{12 \pi k^3 V }}\cos \Big{\{} \frac{\pi}{4}+ \sum_{i=1}^3 \Big{[}\bigl(kj_i+\frac{1}{2}\bigr)\theta_{j_i}+\bigl(kJ_i+\frac{1}{2}\bigr)\theta_{J_i}\Big{]} \Big{\}} \; .
\eea
\end{theorem}

In all sum and products in the sequel the indexes of $v$ run from $1$ to $4$ while those of $p$ run from $1$ to $3$. Thus for example $\prod (t-v_i)$ denotes $\prod_{i=1}^4(t-v_i)$ whereas $\prod (p_j-t)$ denotes $\prod_{j=1}^3(p_j-t)$.

\section{Proof of the main theorem}

Our proof builds on the techniques developed in \cite{Gurau:2005qm}. We start by expressing all factorials in eq. (\ref{Racah}) by Stirling's formula. Step two consists in approximating the discrete sum over $t$ in eq. (\ref{Racah}) by an integral. In step three we give an asymptotic expression for this integral using a saddle point approximation
\footnote{The proofs of \cite{Roberts,Freidel:2002mj} also use saddle point approximations for some integral representations of the $6j$ symbol.}, and obtain theorem \ref{theo}.

As we are interested only in the dominant behavior we will allways consider only first order approximations, so that troughout this paper $=$ will mean equal up to a multiplicative factor $1+1/k$.

\subsection{The prefactor}

We use Stirling's formula
\bea
n!=\sqrt{2\pi}e^{(n+\frac{1}{2})\ln(n)-n}\; ,
\eea
to express all the factorials. Thus, a typical triangle coefficient will be
\bea
&&\Delta(kj_1,kj_2,kj_3)=\frac{2\pi}{[k(j_1+j_2+j_3)+1]} \nonumber\\
&&e^{[k(j_1+j_2-j_3)+\frac{1}{2}]\ln[k(j_1+j_2-j_3)]-k(j_1+j_2-j_3)}\nonumber\\
&&e^{[k(j_1-j_2+j_3)+\frac{1}{2}]\ln[k(j_1-j_2+j_3)]-k(j_1-j_2+j_3)}\nonumber\\
&&e^{[k(-j_1+j_2+j_3)+\frac{1}{2}]\ln[k(-j_1+j_2+j_3)]-k(-j_1+j_2+j_3)}\nonumber\\
&&e^{-[k(j_1+j_2+j_3)+\frac{1}{2}]\ln[k(j_1+j_2+j_3)]+k(j_1+j_2+j_3)}
\; ,
\eea
were we separated the first term in the denominator. A straightforward computation gives
\bea \label{Delta()simplu}
&&\Delta(kj_1,kj_2,kj_3)=2\pi e^{\frac{1}{2}\ln\frac{(j_1+j_2-j_3)(j_1-j_2+j_3)(-j_1+j_2+j_3)}{(j_1+j_2+j_3)^{3}}}
\nonumber\\
&&e^{k[(j_1+j_2-j_3)\ln(j_1+j_2-j_3)+(j_1-j_2+j_3)\ln(j_1-j_2+j_3)+(-j_1+j_2+j_3)\ln(-j_1+j_2+j_3)]}\nonumber\\
&&e^{-k[(j_1+j_2+j_3)\ln(j_1+j_2+j_3)]} \; .
\eea

The prefactor of the sum in eq. (\ref{Racah}) is a product of four such triangle coefficients. After some manipulations it can be put into the form
\bea\label{prefac}
(2\pi)^2 e^{H(j,J)+kh(j,J)} \;,
\eea
with
\bea \label{prefac1}
h(j,J)=&&j_1 h_{j_1} + j_2 h_{j_2}+ j_3 h_{j_3}+ J_1 h_{J_1}+ J_2 h_{J_2}+ J_3 h_{J_3}\nonumber\\
h_{j_1}=&&\frac{1}{2}\ln\Big{\{}\frac{(j_1+j_2-j_3)(j_1-j_2+j_3)}{(j_1+j_2+j_3)(-j_1+j_2+j_3)} \nonumber\\
&&\frac{(j_1+J_2-J_3)(j_1-J_2+J_3)}{(j_1+J_2+J_3)(-j_1+J_2+J_3)} \Big{\}} \; ,
\eea
and
\bea
H(j,J)=\frac{1}{2}(h_{j_1} + h_{j_2}+ h_{j_3}+ h_{J_1}+ h_{J_2}+ h_{J_3}) \; .
\eea
\subsection{The integral approximation}

We now turn our attention to the sum in eq. (\ref{Racah}), which we denote by $\Sigma$. Denoting $v_1=j_1+j_2+j_3$ etc. (hence without the scale factor $k$), and separating the $t+1$ term in the numerator, the sum becomes
\bea
\Sigma=\frac{1}{(2\pi)^3}\sum_{k\text{ max }v_i}^{k \text{ min }p_j} e^{g(t)} \;,
\eea
with
\bea
&&g(t)=\imath \pi t+\ln(t+1)+\frac{1}{2}\ln\frac{t}{\prod(t-kv_i)\prod(kp_j-t)}+t[\ln(t)-1]\nonumber\\
&&-\sum(t-kv_i)[\ln(t-kv_i)-1]-\sum(kp_j-t)[\ln(kp_j-t)-1] \nonumber\\
&&=\imath \pi t+\ln(t+1)+\frac{1}{2}\ln\frac{t}{\prod(t-kv_i)\prod(kp_j-t)}\nonumber\\
&&+t\ln(t)-\sum(t-kv_i)\ln(t-kv_i)-\sum(kp_j-t)\ln(kp_j-t) \;, 
\eea
where in the last equality we have used $\sum_i v_i=\sum_j p_j$.

We change variables to $t= k x$. The exponent rewrites as
\bea\label{exponent}
&&\frac{1}{2}\ln\frac{x^3}{k^4\prod(x-v_i)\prod(p_j-x)}+k\Big{\{} \imath \pi x+x\ln(x)\nonumber\\ 
&&-\sum(x-v_i)\ln(x-v_i)-\sum(p_j-x)\ln(p_j-x) \Big{\}}\;.
\eea
Taking out the $k$ in the first logarithm we write the sum as
\bea \label{sum}
\Sigma=
\frac{1}{(2\pi)^3} \sum_{x=\text{ max }v_i}^{\text{min }p_j} \frac{1}{k^2} e^{F(x)+kf(x)} \; ,
\eea
where $F(x)$ and $f(x)$ can be read out of eq. (\ref{exponent}).

The sum (\ref{sum}) is identified as a Riemann sum. We approximate it by an integral and taking into account that one $k^{-1}$ factor in eq. (\ref {sum}) 
plays the role of $dx$ we have
\bea\label{integral}
\Sigma=
\frac{1}{(2\pi)^3 k} \int_{\text{ max }v_i}^{\text{min }p_j} dx~e^{F(x)+kf(x)} \; .
\eea

Note that eq. (\ref{Delta()simplu}) and (\ref{integral}) can be used to defined a $\{6j\}$ symbol with not only integer and halfinteger entries, but also continuous entries. This continuous version is an analytic continuation of the symbol with (half-)integer entries.

As $k$ is a large parameter the integral (\ref{integral}) can be computes by a saddle point approximation.
Taking into account eq. (\ref{prefac}) we find the following contribution of a saddle point $x_s$ to the value of the $6j$ symbol
\bea \label{truc}
\frac{1}{\sqrt{2 \pi k^3}} \frac{1}{\sqrt{-f''(x_s)}}e^{H(j,J)+F(x_s)+k[h(j,J)+f(x_s)]} \, .
\eea

\subsection{The Saddle Points}

The saddle points equation is 
\bea
f'(x)=\imath \pi+\ln(x)-\sum \ln(x-v_i)+\sum \ln(p_j-x)=0 \; ,
\eea
that is
\bea \label{eq:saddle}
x(p_1-x)(p_2-x)(p_3-x)=-(x-v_1)(x-v_2)(x-v_3)(x-v_4).
\eea
The coefficients of $x^4$ and $x^3$ compute to zero. The saddle point equation becomes
\bea
Ax^2-Bx+C=0 \; ,
\eea
with
\bea
A&=&-\sum_{k<l} p_kp_l+\sum_{i<j} v_i v_j=2(j_1J_1+j_2J_2+j_3J_3)\nonumber\\
B&=&-p_1p_2p_3+\sum_{i<j<k} v_iv_jv_k\nonumber\\
&=&2\big{[}(j_1J_1+j_2J_2+j_3J_3)(j_1+J_1+j_2+J_2+j_3+J_3)\nonumber\\
&+&j_1j_2j_3+J_1j_2J_3+J_1J_2j_3+j_1J_2J_3
\big{]}\nonumber\\
C&=&v_1v_2v_3v_4 \; .
\eea

To solve this equation start by computing its discriminant (denoted by $\Delta$ with no arguments)
\bea
\frac{4AC-B}{4}=\frac{\Delta}{4}&=&j_1^2J_1^2(j_2^2+J_2^2+j_3^2+J_3^2-j_1^2-J_1^2)\nonumber\\
&+&j_2^2J_2^2(j_1^2+J_1^2+j_3^2+J_3^2-j_2^2-J_2^2)\nonumber\\
&+&j_3^2J_3^2(j_2^2+J_2^2+j_1^2+J_1^2-j_3^2-J_3^2)\nonumber\\
&-&j_1^2j_2^2j_3^2-J_1^2j_2^2J_3^2-J_1^2J_2^2j_3^2-j_1^2J_2^2J_3^2 \; .
\eea
Substituting all $j$'s and $J$'s in terms of $j_1,J_2,j_3$ using eq. (\ref{mese}) we find
\bea
\frac{\Delta}{4^2}=j_1^2[\vec{J}_2\wedge \vec{j}_3]^2-[\vec{j}_1\wedge(\vec{J}_2\wedge \vec{J}_3)]^2=
[\vec{j}_1\cdot(\vec{J}_2\wedge \vec{j}_3)]^2=6^2 V^2 \; ,
\eea
where $V$ is the volume of the tetrahedron $j_1,j_2,j_3,J_1,J_2,J_3$!

The two saddle points are then
\bea\label{eq:sol}
x_{\pm}=\frac{B\pm \imath\sqrt{\Delta}}{2A} \; .
\eea

\subsection{The contributions of the saddle points}

We use eq. (\ref{exponent}) to express $f$ as
\bea \label{func}
f(x)=x\ln\Big{(}
\frac{-x\prod(p_j-x)}{\prod(x-v_i)}
\Big{)}+\sum v_i\ln(x-v_i)-\sum p_j\ln(p_j-x) \ .
\eea

Using eq. (\ref{eq:saddle}) we see that, at a saddle point, the first term above is zero. Hence
\bea
f(x_{\pm})=\sum v_i\ln\Big{(}\frac{B}{2A}-v_i\pm \imath \frac{\sqrt{\Delta}}{2A}\Big{)}
-\sum p_j \ln\Big{(}p_j-\frac{B}{2A}\mp \imath \frac{\sqrt{\Delta}}{2A}\Big{)} \; .
\eea

The real part of $f$ is equal for the two saddle points, hence both are dominant. It is then necessary to take the sum of the two contributions.

We analyze the contribution of $t_+$. Substituting (\ref{eq:vp}) in (\ref{func}), $f(t_+)$ writes as a sum over the six numbers $j$ and $J$
\bea
f(t_+)=j_1 f_{j_1}+j_2 f_{j_2}+j_3 f_{j_3}+J_1 f_{J_1}+J_2 f_{J_2}+J_3 f_{J_3} \; , 
\eea
where 
\bea \label{eq:fj}
f_{j_1}= \ln \Big{[}\frac{(x_+-v_1)(x_+-v_4)} {(p_2-x_+)(p_3-x_+)} \Big{]} \; .
\eea

\subsection{The second derivative}

The second derivative will give a volume factor and an extra piece which we will combine with the $F(x_+)$ term in the exponential. 
We start by computing the second derivative at the saddle point $x_+$
\bea
-f''(x_+)=\sum \frac{1}{x_+-v_i}+\sum \frac{1}{p_j-x_+}-\frac{1}{x_+} \; .
\eea
Using the saddle point equation (\ref{eq:saddle}) we have
\bea
\frac{1}{x_+-v_1}=\frac{-(x_+-v_2)(x_+-v_3)(x_+-v_4)}{x_+\prod (p_j-x_+)} \; ,
\eea
and substituting the first four factors yields
\bea
&&-f''(x_+)=\frac{1}{x_+\prod (p_j-x_+)}\nonumber\\
&&\Big{[}
-[(x_+-v_2)(x_+-v_3)(x_+-v_4)+(x_+-v_1)(x_+-v_3)(x_+-v_4) \nonumber\\
&&+(x_+-v_1)(x_+-v_2)(x_+-v_4) +(x_+-v_1)(x_+-v_2)(x_+-v_3)]\nonumber\\
&&+x_+[(p_1-x_+)(p_2-x_+)+(p_1-x_+)(p_3-x_+)+(p_2-x_+)(p_3-x_+)]\nonumber\\
&&-(p_1-x_+)(p_2-x_+)(p_3-x_+)] 
\Big{]} \; .
\eea
The numerator of the above fraction computes to
\bea
&&x_+(-2\sum_{i<j}v_iv_j+2\sum_{k<l}p_kp_l)+(\sum_{i<j<k}v_iv_jv_k-p_1p_2p_3)\nonumber\\
&&=-2x_+A+B=-\imath\sqrt{\Delta} \; .
\eea

Hence 
\bea\label{deriv2}
-f''(x_+)=\frac{-\imath\sqrt{\Delta}}{x_+\prod (p_j-x_+)} \; .
\eea

Substituting eq. (\ref{deriv2}) into (\ref{truc}) gives
\bea
\frac{1}{\sqrt{2\pi k^3 (-\imath) \sqrt{\Delta}}}e^{H+\frac{1}{2}\ln[x_+\prod(p_j-x_+)]+F(x_+)+k[h+f(x_+)]} \; .
\eea
Putting together $F(x_+)$ and the contribution given by $f''(x_+)$ we get  
\bea
&&\frac{1}{2}\ln[x_+\prod(p_j-x_+)]+F(x_+)=\frac{1}{2}\ln\frac{x_+^4}{\prod(x_+-v)}\nonumber\\
&&=\frac{1}{2}\ln\frac{\prod(x_+-v)^3}{\prod(p-x_+)^4}=\frac{1}{2}(f_{j_1}+f_{j_2}+f_{j_3}+f_{J_1}+f_{J_2}+f_{J_3}) \; .
\eea

We conclude that the contribution of the $x_+$ saddel point is
\bea\label{eq:contrib}
\frac{1}{\sqrt{2 \pi k^3(-\imath) \sqrt{\Delta}}} 
&&e^{(k j_1+\frac{1}{2})(h_{j_1}+f_{j_1})+(k j_2+\frac{1}{2})(h_{j_2}+f_{j_2})+(k j_3+\frac{1}{2})(h_{j_3}+f_{j_3})}\nonumber\\
&&e^{(k J_1+\frac{1}{2})(h_{J_1}+f_{J_1})+(k J_2+\frac{1}{2})(h_{J_2}+f_{J_2})+(k J_3+\frac{1}{2})(h_{J_3}+f_{J_3})} \, .
\eea

\subsection{Final evaluation}
We must compute $f_j$. We use eq. (\ref{eq:fj}) and compute separately the real and imaginary part. The imaginary part is
\bea
\Im (f_{j_1})=\theta_{j_1}&=&\text{Arg }(t_+-v_1)+\text{Arg }(t_+-v_4)\nonumber\\
&+&\text{Arg }(p_2-t_-)+\text{Arg }(p_3-t_-) \; .
\eea
Using eq. (\ref{eq:sol}) we write the four arguments in the above equation as
\bea
&&\text{Arg }(t_+-v_1)=\text{Atan}\Big{(}\frac{\sqrt{\Delta}}{B-2Av_1}\Big{)} \nonumber\\ 
&&\text{Arg }(t_+-v_4)= \text{Atan}\Big{(}\frac{\sqrt{\Delta}}{B-2Av_4}\Big{)}\nonumber\\
&&\text{Arg }(p_2-t_-)= \text{Atan}\Big{(}\frac{\sqrt{\Delta}}{2Ap_2-B}\Big{)}\nonumber\\
&&\text{Arg }(p_3-t_-)=\text{Atan}\Big{(}\frac{\sqrt{\Delta}}{2Ap_3-B}\Big{)} \; .
\eea
Taking into account that
\bea
&&\tan(a_1+a_2+a_3+a_4)\nonumber\\
&&=\frac{\sum_i\tan(a_i)-\sum_{i<j<k}\tan(a_i)\tan(a_j)\tan(a_k)}
{1-\sum_{i<j}\tan(a_i)\tan(a_j)+\tan(a_1)\tan(a_2)\tan(a_3)\tan(a_4)} \; ,
\eea
a straightforward but extremely tedious computation shows that
\bea
&&\tan (\theta_{j_1})=\nonumber\\
&&\frac{j_1\sqrt{\Delta}}{j_1^2(j_1^2+2J_1^2-j_2^2-J_2^2-j_3^2-J_3^2)+j_2^2J_3^2+j_3^2J_2^2-j_2^2J_2^2-j_3^2J_3^2} \; .
\eea
Substituting again $j_2, J_3$ and $J_1$ using eq. (\ref{mese}) we find
\bea \label{eq:im}
\tan(\theta_{j_1})=\frac{4 j_1 [\vec{j_1}\cdot (\vec{J_2}\wedge\vec{j_3})]}
{4 (\vec{j_1}\wedge\vec{j_3})\cdot(\vec{J_2}\wedge \vec{j_1})}=
\frac{|(\vec{J_2}\wedge \vec{j_1})\wedge(\vec{j_1}\wedge\vec{j_3})|}
{(\vec{J_2}\wedge \vec{j_1})\cdot(\vec{j_1}\wedge\vec{j_3})} \; .
\eea
As the vectors $\vec{J_2}\wedge \vec{j_1}$ and $\vec{j_1}\wedge \vec{j_3}$ are normal (and outward pointing) to the planes
$j_1, J_2, J_3$ and $j_1,j_2,j_3$ we identify $\theta_{j_1}$ as the (exterior) dihedral angle of the tetrahedron.

We now turn our attention to the real part of $f_{j_1}$
\bea
\Re(f_{j_1})&=&\ln\Big{|}\frac{(t_+-v_1)(t_+-v_4)} {(p_2-t_+)(p_3-t_+)} \Big{|}\nonumber\\
&=&
\frac{1}{2}\ln\Big{[}\frac{[(B-2Av_1)^2+\Delta][(B-2Av_4)^2+\Delta]}{[(2Ap_2-B)^2+\Delta][(2Ap_3-B)^2+\Delta]}\Big{]}
\nonumber\\
&&=\frac{1}{2}\ln\frac{(Av_1^2-Bv_1+C)(Av_4^2-Bv_4+C)}{(Ap_2^2-Bp_2+C)(Ap_3^2-Bp_3+C)} \; .
\eea
Again a straightforward but tedious computation shows the real part equals
\bea
\frac{1}{2}\ln\frac{(j_1+J_2+J_3)(-j_1+J_2+J_3)(j_1+j_2+j_3)(-j_1+j_2+j_3)}
{(j_1+J_2-J_3)(j_1-J_2+J_3)(j_1+j_2-j_3)(j_1-j_2+j_3)} \; ,
\eea
and using eq. (\ref{prefac1}) we conclude that
\bea\label{eq:re}
h_{j_1}+\Re( f_{j_1})=0 \; .
\eea

Collecting eq. (\ref{eq:contrib}), (\ref{eq:im}) and (\ref{eq:re}) yields the following contribution of the $x_+$ saddle point
\bea
\frac{1}{\sqrt{48 \pi k^3 V }}e^{\imath \frac{\pi}{4}+\imath \sum_{i=1}^3\bigl[\bigl( kj_i+\frac{1}{2}\bigr) \theta_{j_i}+ \bigl( kJ_i+ \frac{1}{2}\bigr) \theta_{J_i}\bigr]} \; .
\eea

Summing the constributions of $x_+$ and $x_-$ proves Theorem \ref{theo}.

\section{Conclusion}

Our proof is easily adapted to the Minkowskian tetrahedron. In that case the discriminant changes sign (as the volume becomes imaginary). We find two real saddle points, and only one of the two is dominant. The computations are essentially the same, and it is easy to recover the expected exponential decay.

This method can be genralized to higher $3nj$'s symbols. One needs first to rexpress them as multiple sums and the proceed in a parallel way. For the $9j$ symbol, for instance one should use the three sums formula \cite{trip}. The saddle point equations become more involved but the computations should be manageable.

\section{Acknowledgements}

Research at Perimeter Institute is supported by the Government of Canada through Industry 
Canada and by the Province of Ontario through the Ministy of Research and Innovation.

\end{document}